\input psfig
\documentstyle[twocolumn,aps]{revtex}
\begin{document}
\draft
\twocolumn[\hsize\textwidth\columnwidth\hsize\csname @twocolumnfalse\endcsname
%
%

\title{ Influence of Hole Doping on
Antiferromagnetic Real-Space Approaches for the High-Tc Cuprates }

\author{ Alexander Nazarenko$^1$, Stephan Haas$^2$, Jose Riera$^3$, 
 Adriana Moreo$^1$,  and Elbio Dagotto$^1$}

\address{$1.$ Department of Physics, and National High Magnetic Field Lab,
Florida State University, Tallahassee, FL 32306, USA}
\address{$2.$ Theoretische Physik, Eidgen\"ossische Technische
Hochschule, 8093 Z\"urich, Switzerland}
\address{$3.$ Instituto de Fisica Rosario, Avda. 27 de Febrero 210 bis,
2000 Rosario, Argentina}

\maketitle

\begin{abstract} Recently proposed scenarios for the cuprates make
extensive use of a ``flat'' quasiparticle (q.p.) dispersion and short-range
hole-hole interactions in real-space,  
both caused by antiferromagnetic (AF) correlations. The density
of states (DOS) at half-filling has a robust
peak which boosts the superconducting critical temperature $T_c$ to
large values as holes are introduced into the (rigid) q.p. band.
Here, the
stability of such scenarios is studied after a 
$finite$ but small 
hole density is introduced. The overall conclusion is that the
main features of real-space AF-based approaches
remain qualitatively similar, namely a large  $T_c$ is found
and superconductivity (SC) appears in the
${\rm d_{x^2 - y^2}}$ channel. As the hole density
grows the chemical potential $\mu$ crosses a broad peak in the DOS.
We also observe that extended 
s-wave SC competes
with d-wave in the overdoped regime.

\end{abstract}

\pacs{74.20.-z, 74.20.Mn, 74.25.Dw}
]
\narrowtext

%
%


The presence  of ``flat'' regions in the normal-state q.p. 
dispersion is a remarkable feature of the phenomenology of hole-doped
high temperature superconductors.\cite{flat-exper}
These flat
bands are located around momenta
${\bf p} = (\pi,0)$ and $(0,\pi)$, and at optimal doping they are
${\rm \sim 10meV}$ below the Fermi energy, according to
angle-resolved photoemission (ARPES) studies.\cite{flat-exper} 
Antiferromagnetic correlations may play
an important role in the generation of these features, as has been
suggested by studies of holes in 
the 2D t-J and Hubbard models at and away 
from half-filling.\cite{flat,bulut2,hanke} In addition, the
flat regions induce a large peak in the DOS which can be used to 
boost $T_c$. This leads to a natural explanation for the existence of an
``optimal doping''  which in this framework 
occurs when the flat regions are reached by $\mu$. 
We will refer to these ideas as the ``Antiferromagnetic van
Hove'' (AFVH) scenario.\cite{afvh} 
Superconductivity in the ${\rm d_{x^2 - y^2}}$ channel is natural in the
AFVH scenario due to the strong AF correlations.\cite{doug,review}
The interaction of holes is better visualized in
$real$ $space$ i.e. with pairing occuring when dressed holes
share a  spin polaronic cloud, as in the spin-bag mechanism.\cite{bob}
This real-space picture (see also Ref.\cite{joynt}) 
holds even for a small AF
correlation length, $\xi_{AF}$, and in this approach
there is no need to tune
parameters to work very close to an AF instability.
Previous scenarios have also used van Hove (vH) singularities 
in the band structure to increase
$T_c$,\cite{vh} but d-wave SC is not natural in this context unless 
AF correlations are included.

%
%




To obtain quantitative information from these intuitive ideas, 
holes moving with a dispersion calculated 
using one hole in an AF background, $\epsilon_{AF}({\bf p})$,
and interacting through a nearest-neighbors (NN) attractive potential,
that mimics the sharing of spin polarons, have been previously
analyzed.\cite{afvh}
Within a rigid band
filling of $\epsilon_{AF}({\bf p})$ and using a BCS formalism,
${\rm d_{x^2 - y^2}}$ superconductivity dominates with
$T_c \sim 100K$ caused by the large
DOS implicit in the hole dispersion.
However, the accuracy of the rigid band approximation has not been
studied before. In particular,
the following questions arise:
(i) does the q.p. peak in the DOS found at $\langle n \rangle =1$
survive a finite hole density?;
(ii) how does the q.p. dispersion change with hole doping, and to what
extend does this affect previous calculations in this framework?;
(iii) are the ``shadow'' regions generated by
AF correlations\cite{bob,shadow} important for
real-space pairing approaches? In this paper all these issues are
discussed. The overall conclusion is that as long as the
hole density is such that the 
$\xi_{AF}$ is at least of a couple of lattice spacings, the 
predictions of the original AFVH scenario\cite{afvh} and other similar
theories remain qualitatively the same.


To study the presence of a large DOS peak away from half-filling,
$N(\omega) (= \sum_{\bf p} A({\bf p},\omega) )$ has been calculated numerically
in systems with strong AF correlations.
While accurate extrapolations to the bulk limit are difficult,
the simple qualitative picture emerging from these studies seems robust.
In Fig.1a, $N(\omega)$ for the 2D t-J model obtained with exact
diagonalization (ED) techniques is shown at several densities.\cite{review} At
half-filling, a large DOS peak caused by
flat regions in the q.p. dispersion appears at the top of the valence 
band, as discussed before.\cite{flat} Weight exists at energies
far from $\mu$ i.e. the large peak carries only part 
of the total weight. As shown in Ref.\cite{flat} the maximum
in the DOS is $not$ reached at the top of the valence band but at slightly
smaller energies. As
$\langle n \rangle$ decreases, the peak is now much 
broader but it remains well-defined. 
At $\langle n \rangle = 0.87$, $\mu$ 
is located close to the energy where
$N(\omega)$ is maximized.
If a source of hole attraction exists in this system,
a SC gap would open at $\mu$, and  
the resulting $T_c$ could be large. However, even with 
$T_c \sim 100K$
the SC gap would not alter
much the already robust peak in the DOS shown in Fig.1a.
For $\langle n \rangle \sim 0.75$, $\mu$ moves to the left
of the peak, where $T_c$ should become smaller. Then, as $\langle n \rangle$
is reduced away from half-filling in the 2D t-J model Fig.1a suggests
that $\mu$ travels across a broad DOS peak providing a natural definition
for the ``underdoped, optimal
and overdoped'' densities.\cite{prelovsek} 
Note that since the peak width 
increases substantially with hole density, strictly speaking
the rigid band filling is invalid. However, such an approximation may
have captured the qualitative physics of the problem,
as discussed in this paper, since the DOS peak is not washed out by
a finite hole density.


\begin{figure}[htbp]
\centerline{\psfig{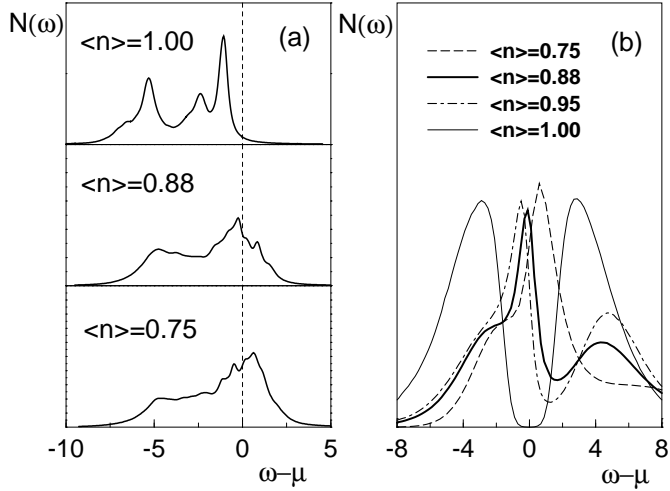}
}
\vspace{.2cm}
\caption{(a) $N(\omega)$ for the 2D t-J model averaging results for 
clusters with
16 and 18 sites, at J/t = 0.4 and for the densities indicated.
The $\delta$-functions were given a width $\eta = 0.25t$. Similar
results were found at other values of J/t;
(b) $N(\omega)$ for the one band Hubbard model obtained on a 
$4 \times 4$ cluster
using QMC and ME. The temperature
is $T=t/4$ and $U/t=12$. 
Densities are indicated. Previous
studies showed that QMC/ME results are only weakly affected by size
effects.[3]
}
\end{figure}

Quantum Monte Carlo (QMC) data for the 2D one-band Hubbard model in
strong coupling 
indicate a qualitative behavior similar to that of the t-J model, i.e.
$\mu$ also crosses a peak as $\langle n \rangle$ varies. This
effect was observed in previous simulations,\cite{bulut1} and 
also in our own studies (Fig.1b).\cite{comm2} 
However, note that
QMC simulations must be supplemented by a Maximum
Entropy (ME) analysis to obtain $N(\omega)$.
ME predictions are more accurate close to $\omega \approx 0$ than at finite
frequency, and thus we here do $not$ interpret Fig.1b as suggesting the
 ``generation'' by doping of a Kondo-like
peak at the top of the valence band
which does not exist at $\langle
n \rangle =1$. We believe that 
the absence of a sharp peak at half-filling is caused by
the systematic errors of the ME procedure and the finite temperature of
the simulation. Actually, a robust peak at $\langle n \rangle =1$
appears in other computational studies,\cite{flat,bulut1}
and in the t-J model (Fig.1a). 
Experimentally in the cuprates it has
been already established\cite{fujimori}  that the states observed  in PES
upon doping are already present in the insulator and are $not$ Kondo
resonances. In the present calculation, Fig.1a suggests that
the broad peak observed at $\langle n \rangle < 1$ in
the t-J model is smoothly connected 
to the sharper peak found at half-filling.




Similar conclusions can be reached
in geometries other than the 2D square lattice. 
In particular, we studied the DOS of the t-J model on a $ladder$,
finding results very similar to those of Fig.1a, in agreement with
previous calculations.\cite{troyer} Note
that on ladders the presence of pairing
at finite hole density has been predicted.\cite{science} 
Thus, a gap may
open at $\mu$ once couplings and lattice sizes
are reached where pairing effects are important.
We have observed such effects in our studies, but we also
found that at $\langle n \rangle = 1$ the DOS has a sharp peak, 
as in  Fig.1a. Then, the ladder DOS peak is
not exclusively caused by
a BCS-induced redistribution of spectral weight upon doping.

In all cases discussed here, calculations of the 
spin-spin correlations show that 
$\xi_{AF}$  is approximately a couple of lattice
spacings when $\mu$ is located near the DOS peak, 
becoming negligible as the overdoped regime is reached. 
Thus, a nonzero $\xi_{AF}$
and $\mu$ near a large DOS peak are $correlated$ features.
It is in this respect that scenarios where AF correlations produce
a large DOS that enhances $T_c$ differ from
vH theories where divergences in the DOS are caused by band
effects already present before interactions are switch on.

%



While the existence of a robust peak in the DOS is in good agreement
with ARPES data,\cite{flat-exper} it is in apparent disagreement with 
specific heat 
studies for YBCO which have been interpreted as corresponding to
a flat DOS.\cite{loram}
The lack of ${\bf p}$-resolution in the
specific heat measurements may solve this puzzle. Actually,
angle-$integrated$ PES results for Bi2212 do not show the sharp flat
features found in ARPES for the same material.\cite{fujimori,imer}
Similar effects may affect the specific heat data.

Now let us discuss the ${\bf p}$-dependence
of the q.p. band obtained from $A({\bf p},\omega)$. 
Results are already available in
the literature.\cite{bulut2,moreo}
In Fig.2a-b, the q.p. dispersion 
is shown at $\langle n \rangle =1$, and at
$\langle n \rangle  \sim 0.87$ for the t-J
model.\cite{moreo} 
Note that upon doping the 
small q.p. bandwidth and the flat regions remain well-defined, inducing a
large DOS peak (Figs.1a).
However, the region around ${\bf p} = (\pi, \pi)$ has changed
substantially, i.e. the AF shadow region clearly observed at
$\langle n \rangle =1$
has reduced its intensity and 
considerable weight has been transferred to the inverse PES region (Fig.2b).
Actually, the q.p. dispersion at $\langle n \rangle \sim 0.87$ 
can be fit by a tight-binding 
nearest-neighbors (NN) dispersion with a small effective hopping
likely associated with $J$. This effect was also noticed in studies of
the Hubbard model.\cite{ortolani,bulut2,hanke}.

The changes in the q.p. dispersion
with hole doping can be interpreted in two ways: First, note that
$A({\bf p}, \omega)$ is influenced by matrix elements of 
$bare$ fermionic 
operators connecting states with $N$ and $N\pm 1$ particles. This 
is important when the q.p. weight is small
i.e. when the state $c_{{\bf p} \sigma} |gs \rangle_N$ does not have a
large overlap with
the ground state of the $N-1$ particles subspace $|gs \rangle_{N - 1}$ 
($|gs \rangle_N$
being the ground state with $N$ particles). It has been
suggested\cite{qp,eder} that if
the hole excitation is created by a  new operator $\gamma_{{\bf p}\sigma}$
that incorporates the dressing of the hole by spin fluctuations,
then $\gamma_{{\bf p}\sigma} |gs \rangle_N$ may now have
a large overlap with $|gs \rangle_{N-1}$.
In other words, if the dressed hole state
resembles an extended spin polaron, then the physics deduced from PES
studies, which rely on the sudden removal of a bare electron from the
system, may be misleading. To the extent that spin polarons remain
well-defined at finite density, the use of $\gamma_{{\bf
p}\sigma}$ will induce spectral weight rearrangements
between the PES and IPES regions, and 
Fig.2b could resemble  Fig.2a after such spectral weight redistribution
takes place, as some results already indicate.\cite{eder}
If this idea were correct, then the study of SC
and transport, both regulated by $dressed$ quasiparticles, could
indeed be handled by filling 
a rigid band given by $\epsilon_{AF}({\bf p})$.

\begin{figure}[htbp]
\centerline{\psfig{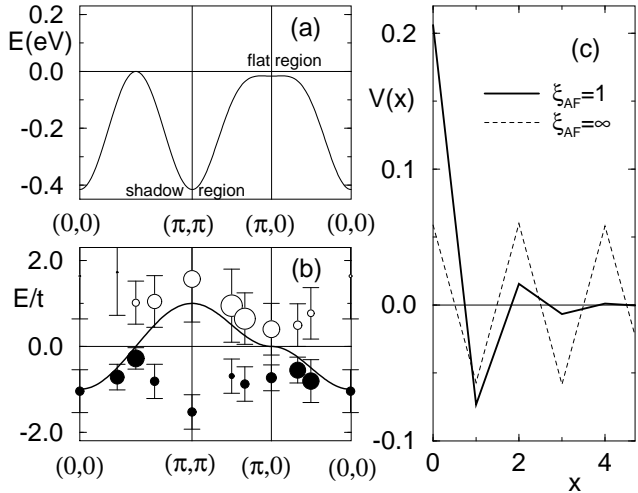}
}
\vspace{.2cm}
\caption{(a) q.p. energy vs momentum obtained at half-filling.[2]
The result shown, $\epsilon_{AF} ({\bf p})$, 
is a good fit of Monte Carlo data on a $12
\times 12$ cluster at $J/t = 0.4$; (b) q.p. dispersion
vs momentum at $\langle n \rangle
= 0.87$ and $J/t = 0.4$ using exact diagonalization of
16 and 18 sites clusters (from Ref.[19]). The open (full)
circles are IPES (PES) results.
Their size is proportional to the peak intensity. The solid line is
the fit $\epsilon_{NN}({\bf p})$;
(d) $V(x)$ along the x-axis after
Fourier transforming the smeared potential $V({\bf p}) = \delta({\bf p} - {\bf
Q})$ (see text). $\xi_{AF}$ is given in lattice units.
}
\end{figure}

However, an alternative is that
the difference between Figs.2a and 2b 
corresponds to an intrinsic change in the q.p. dispersion as $\xi_{AF}$
decreases. This is the less favorable case for the AFVH approach,
and thus we analyze this
possibility in detail here. For this purpose,
we have applied the standard BCS formalism
to a model with a low density of q.p.'s having a dispersion
$\epsilon_{NN}({\bf p})/eV = -0.2(cosp_x + cosp_y)$, which roughly
reproduces the dominant features of Fig.2b. As before, we include
NN attraction between q.p.'s. While naively it may seem
dubious to use the same interaction both at and away from
half-filling, the hole-hole potential should not be much
affected at
distances shorter than $\xi_{AF}$. This can be illustrated by a
real space analysis (Fig.2c) of a smeared $\delta$-function potential
of AF origin $V({\bf q}) = 
\xi_{AF}/ [ 1 + \xi_{AF}^2 ( {\bf q} - {\bf Q} )^2 ]$,
where the lattice spacing is set to 1, and ${\bf Q} = (\pi,\pi)$.
Fig.2c shows that the NN potential ($x=1$)
does not change noticeably as $\xi_{AF}$ is reduced, while $V(x>1)$
is rapidly suppressed. 
Then, using the same
NN form of the potential for many densities
is justified. Note also that the local character of the real-space 
potential does $not$ imply small Cooper pairs. Their size
can be regulated using both its range $and$ intensity.


\begin{figure}[htbp]
\centerline{\psfig{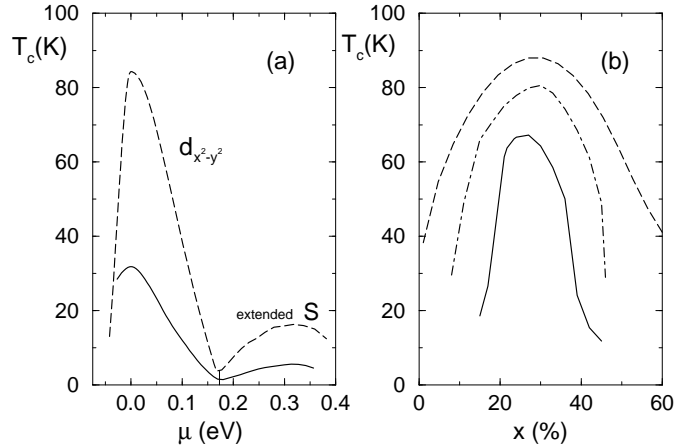}
}
\vspace{.2cm}
\caption{(a) $T_c$ vs $\mu$ obtained with
the BCS gap equation, with $\mu=0$ as the chemical
potential corresponding to the saddle-point. 
The dashed line corresponds to $\epsilon_{AF}({\bf p})$, and the
solid line to $\epsilon_{NN}({\bf p})$. Interpolations between these two
extreme cases can be easily constructed using two band models with
weights regulated by $Z$-factors. Note the presence of 
both ${\rm d_{x^2 - y^2}}$-wave 
and extended s-wave SC; (b) $T_c$ for d-wave SC vs the 
percentage $x$ of filling of
the q.p. band ($not$ of the full hole spectrum).
The dashed line is the same as in (a). The solid
line corresponds to the BCS gap equation result
making zero the weight $Z_{\bf p}$ of
states in the q.p. dispersion that are at energies from the saddle point
larger than 2.5\% of the total bandwidth
(i.e. basically including states only in a window
of energy  $\sim 125K$ around the  flat regions). The
dot-dashed line is the same but using a window of $\sim 250K$ around
the flat regions.
}
\end{figure}

Then, let us discuss the results obtained using the BCS formalism with an
attractive NN potential $V = -0.6J$,\cite{afvh} with $J = 0.125 eV$, for both
dispersions $\epsilon_{AF}({\bf p})$ and 
$\epsilon_{NN}({\bf p})$.
Solving numerically the gap equation,  $T_c$ is shown
in Fig.3a. As reported before,\cite{afvh} SC at
$T_c \sim 80-100K$ in the ${\rm d_{x^2 - y^2}}$ channel 
appears naturally if the AF-dispersion is used. 
If, instead, $\epsilon_{NN}({\bf p})$
is used, the flat bands present in this narrow dispersion produce
$T_c \sim 30K$ which is still large. Even more remarkable is the fact 
that the ${\rm d_{x^2 - y^2}}$ character of the SC state 
is maintained. This result can be understood noticing that a combination of
$\epsilon_{NN} ( {\bf p})$ with
an attractive NN potential effectively locates us in the 
family of ``t-U-V'' models with U repulsive and V attractive, where
it is known that for a ``half-filled'' electronic band the dominant SC
state is ${\rm d_{x^2 - y^2}}$-wave.\cite{micnas} 
In other words, when $\mu$ is at the flat region
in Fig.2b it approximately
corresponds to a ``half-filled'' q.p. band, leading
to a ${\rm d_{x^2 - y^2} }$-wave SC state (Fig.3a).\cite{edwards} 
Then, one of our main results is that 
even if the q.p. dispersion changes substantially with
doping near the ${\bf Q}$ point, such an effect
does $not$ alter the main qualitative features found in previous
studies.\cite{afvh}

The present analysis also shows that the
AF ``shadow'' regions of  $\epsilon_{AF}({\bf p})$
are not crucial for the success of the real-space approach.
Actually, using $\epsilon_{NN}({\bf p})$, which
does not contain weight in PES near $(\pi,\pi)$, 
$T_c$ is still robust and the d-wave
state remains stable. To make this point more clear, we analyzed $T_c$
using $\epsilon_{AF}({\bf p})$ 
but modulating the contribution
of each momentum with a ${\bf p}$-dependent weight $Z_{\bf p}$ in the
one particle Green's function.
We considered the special case where $Z_{\bf p}$ is
zero away from a window of total width $W$ centered at the
saddle point, which is located in the flat
bands region. Inside the window $W$, $Z_{\bf p} =1$.
Such a calculation also addresses indirectly possible concerns
associated with widths $\sim (\epsilon({\bf p}) -\epsilon_F)^2$
that q.p. peaks would acquire away from half-filling in standard Fermi
liquids. Results are shown in Fig.3b, for d-wave SC.
Note that even in the case where $W$ is as small as just 5\% of the
total bandwidth (itself already small of order $2J$), 
$T_c$ remains robust and close to $70K$. Then, it
is clear that the dominant contribution to $T_c$ comes from the flat regions
and the shape of the q.p. dispersion away from them is
not relevant for the sucess of the real-space approach.



We end this paper with novel predictions obtained from
$\epsilon_{AF}({\bf p})$ 
and $\epsilon_{NN}({\bf p})$, when $\mu$ reaches the
bottom of these bands (which should $not$ be confused with 
the bottom of the full hole spectrum since a
large amount of weight lies at energies lower than those of 
the q.p. band). In this regime, the BCS analysis
applied to any of the two dispersions
shows that extended $s$-wave SC dominates over 
${\rm d_{x^2 - y^2}}$-wave SC in the ``overdoped''
regime (Fig.3a). This occurs when the q.p. band is
nearly empty which, according to Fig.1a, corresponds to an overall
density of $\langle n \rangle \sim 0.7$. 
This change in the symmetry of the SC state
can be understood recalling again that $\epsilon_{NN}({\bf p})$.
supplemented by a NN attraction
formally corresponds to an effective ``t-U-V'' model. It is 
well-known that in this model the
SC state symmetry changes from d- to s-wave as the
density is reduced away from half-filling to a nearly empty
system.\cite{micnas,dago5}
Actually the bound state of two particles with a NN tight binding
dispersion and NN attraction is $s-wave$.
To the extend that the AF- or NN-dispersions survive up to 
$\sim 25\%$ hole doping, as suggested by numerical data,\cite{bulut1,moreo}
scenarios based on the real-space interaction of q.p.'s
predict a competition between 
extended s-wave and ${\rm d_{x^2 - y^2}}$-wave 
SC in the $overdoped$ regime.
It is remarkable that calculations
based on the analysis of the low electronic density $\langle n
\rangle \ll 1$ limit of the t-J model lead to analogous
conclusions.\cite{kolte,dago5} 
Our approach is based on
a very different formalism, but it arrives to similar results,
and thus we believe that a crossover from d-
to s-wave dominated SC in overdoped cuprates could occur.
Recent ARPES data for overdoped Bi2212 
have indeed been interpreted as corresponding to a mixing of s- and d-wave
states.\cite{ma}

Summarizing, the predictions of
real-space theories based on dispersions and interactions
calculated at half-filling are qualitatively stable upon the 
introduction of a finite
hole density. These include a large $T_c$, within the BCS formalism,
and SC in the ${\rm d_{x^2 - y^2}}$ channel. In the overdoped
regime a possible crossover from d- to extended s-wave SC was discussed.


We thank A. Fujimori,
D. Dessau, Z.X. Shen, and D. M. Edwards for useful correspondence.
E. D. and A.M. are supported by grant NSF-DMR-9520776.
We thank the NHMFL and 
MARTECH for additional support.


\end{document}